\documentclass[letterpaper, 10 pt, conference]{ieeeconf}  % Comment this line out if you need a4paper

\IEEEoverridecommandlockouts                             

\usepackage{graphics}
\usepackage[english]{babel}
\usepackage{epsfig} % for postscript graphics files
\usepackage{mathptmx} % assumes new font selection scheme installed
\usepackage{times} % assumes new font selection scheme installed
\usepackage{amsmath,mathtools} 
\allowdisplaybreaks
\usepackage{array}
\usepackage{mdwmath}
\usepackage{amssymb} 

\newtheorem{assumption}{Assumption}
\usepackage{algorithm}
\usepackage{algpseudocode}
\newcommand{\comment}[1]{{ }}
\title{\LARGE \bf
Data-Driven Newton Raphson Controller Based on Koopman Operator Theory
}

\author{Mi Zhou $^{1}$   
\thanks{$^{1}$  Mi Zhou is with the School of Electrical and Computer Engineering, Georgia Institute of Technology, Atlanta, GA 30332.
        {\tt\small mzhou91@gatech.edu}
        }
}

\begin{document}
\maketitle
\thispagestyle{empty}
\pagestyle{empty}
\begin{abstract}
Newton-Raphson controller is a powerful prediction-based variable gain integral controller.
Basically, the classical model-based Newton-Raphson controller requires two elements: the prediction of the system output and the derivative of the predicted output with respect to the control input.
In real applications, the model may not be known and it is infeasible to predict the system sometime ahead and calculate the derivative by finite difference method as done in simulation.
To solve these problems, in this work, we utilize the Koopman operator framework to reconstruct a linear model of the original nonlinear dynamical system and then utilize the output of the new linear system as the predictor of the Newton-Raphson controller.
This method is only based on collected data within some time instant thus more practical.
Three examples related to highly nonlinear systems are provided to verify the effectiveness of our proposed method.
\end{abstract}

\section{INTRODUCTION}
Trajectory tracking control is one of the most important topics in the robotics field such as mobile robots \cite{leaderfor}, self-driving cars \cite{selfdriving}, quadrotor UAVs \cite{UAV}, underwater robots \cite{underwater}, and so on. 
Many methods have been proposed to achieve real-time tracking performance.
Existing techniques include Proportional-Integral-Derivative (PID) \cite{underwater}, Byrnes-Isidori regulator \cite{regulator}, model predictive control \cite{mpcbook,MPC}, etc.

Newton Raphson (NR) controller, first proposed in \cite{Wardi}, is a tracking method that based on variable-gain integrator and the Newton-Raphson method for finding zeros of a function. 
This technique consists of three elements: (i) output prediction which tracks the reference signal; (ii) an integral controller with variable gain; (iii) a speedup of the control action for enhancing the tracker's accuracy and guaranteeing the stability of the closed-loop system.
\cite{Wardidetail} provided a detailed introduction to this technique and theoretical derivation about the convergence of the tracking controller and error analysis with persuasive illustrative simulation and laboratory experiments.
Subsequently, more works appeared that used Newton Raphson controller to solve several challenging problems such as tracking control of leader-follower multi-agent systems \cite{Wardi,leaderfor}, distributed formation control of multi-agent mobile systems in swarms and platoons \cite{formation}, driving an underactuated system in a potentially adversarial environment modeled as a pursuit-evasion game \cite{TA}, tracking control of nonlinear inverted pendulums and differentially driven cars \cite{Niu}, and so on.
All these works showed that the tracking convergence using this method is quantified for both constant or time-dependent reference signals and was quite fast and had a large region of convergence as well.
The NR regulation technique described above is based on a look-ahead simulation of the systems which works as both a predictor and an observer.
This mechanism, however, requires a precise model of the system in order to obtain a reliable output prediction and hence an effective tracking performance.
Therefore, some works started to explore the potential of neural networks in the predictor.
\cite{TA} formulated a pursuit-evasion game, regarded it as a tracking regulation problem solved using the Newton-Raphson controller, and used a deep neural network to approximate the behavior of the evader gathered online during the pursuit.
In \cite{Niu}, the authors utilized a feed-forward neural network as an output predictor for the Newton-Raphson controller thus achieving a model-free realization.
However, the training process has a high reliance on the accuracy of the data thus it lacks robustness and real-time realization.

There is an increasing need for data-driven system identification approaches with the development of more complicated robotic systems.
One alternative to data-driven approaches is to use neural networks and deep learning to identify a system.
However, the deep-learning method suffers from long-time training.
The Koopman operator framework offers new opportunities for the control of nonlinear systems from the perspective of linearizing systems.
It is a super powerful tool for identifying and linearizing a nonlinear system with higher accuracy compared to traditional linearization methods.
As we know, the computational limitation due to nonlinearity is an essential challenge in robotics control.
Instead of linearizing systems directly, Koopman analysis achieves linearization by representing the nonlinear dynamics in a globally linear framework.
The Koopman-based approach is a perfect model-free system identification method with low time complexity.
It has thus found wide applications in model predictive control, active learning, and hybrid control of robots.

There are loads of works related to Koopman operator theory with good theoretical foundations and applications in real robotic systems.
A detailed introduction to Koopman operator theory can be found in \cite{Brunton2019NotesOK}.
In \cite{soft}, the authors used model predictive control (MPC) to control soft continuum manipulator arms after using Koopman operator theory to identify the soft manipulator models.
In \cite{DualModeMPC}, the authors used Koopman eigenfunctions to lift the nonlinear system dynamics
to provide a linear approach for the design of MPC with state and input constraints.
\cite{aerospace} proposed a high-order optimal control strategy implemented in the Koopman operator framework and tested in the Duffing oscillator and Clohessy-Wiltshire problem which models the relative motion between two satellites.
All these works present good and efficient performance of the Koopman operator in identifying systems.
%%%%%%%%%%%%%%%%%%%%%%%%%%%%%%%%%%%%%%%%%%%%%%%

The objective of this paper is thus to propose a real-time data-driven Newton-Raphson controller by using Koopman linearization.
We will test our tracking algorithm on the Van Der Pol system, an overhead crane system, and a differentially driven car system. 
In all experiments, the system has to track a time-varying reference signal.
We then compare the tracking results with the classical model-based Newton-Raphson controller.

This paper is organized as follows: in Section \ref{sec:PS}, we formulate our problem. 
In Section \ref{sec:pc}, the Koopman operator theory and the proposed controller are introduced in detail.
Section \ref{sec:exp} provides three examples to illustrate the efficiency of our controller by comparing it with the classical model-based Newton-Raphson controller.
We finally conclude our article in Section \ref{sec:conclude}.

\section{Problem statement} \label{sec:PS}
Consider the following nonlinear system:
\begin{align} \label{eqn:system}
\dot x (t)= f(x(t),u(t))
\end{align}
with the output equation as
\begin{align}
y(t) = h(x(t))
\end{align}
where $x \in \mathbb{R}^n$, $u(t)\in \mathbb{R}^m$, $f(x(t),u(t))$ is continuously differentiable in $x$ for every $u\in \mathbb{R}^m$, and continuous in $u$ for every $x\in \mathbb{R}^n$, and $h(x(t))$ is continuously differentiable.
Moreover, to make sure Eqn. \eqref{eqn:system} has a unique solution on $t\in [0, \infty)$, we make the following assumptions:
\begin{assumption}[\cite{leaderfor}]
\begin{enumerate}
    \item For every compact set $\Gamma_1 \subset \mathbb{R}^n$ and $\Gamma_2 \subset \mathbb{R}^m$, the functions $f(x(t),u(t))$ and $\frac{\partial f}{\partial x}(x(t),u(t))$ are Lipschitz continuous on $\Gamma_1 \times \Gamma_2$.
    \item For very compact set $\Gamma_2 \subset \mathbb{R}^m$, there exists $K>0$ such that, for every $x\in \mathbb{R}^n$ and $u\in \Gamma_2$,
    \begin{align*}
     ||f(x,u)|| \leq K(||x||+1).
    \end{align*}
\end{enumerate}
\end{assumption}
Define $r(t) \in \mathbb{R}^k$ as the reference signal.
The output tracking control problem is defined as
\begin{align}
\lim_{t\rightarrow \infty} ||r(t)-y(t)|| =0
\end{align}
which can also be viewed as solving the root of time-dependent equations $r(t)-y(t)=0$.
This brings the idea of designing a controller with the following iterative form:
\begin{align}
    u_{n+1} = u_n - \frac{r(t)-y(t)}{(r(t)-y(t))'}
\end{align}
to find the root (i.e., controller) $u(t)$.

In the design of the Newton-Raphson controller, the prediction phase is, saying, at time $t$, we can predict the system from time $t$ to $t+T$ by solving the following differential equation
\begin{align}\label{eqn:deri2}
\dot{\Tilde{x}} (\tau) = f(\Tilde{x}(\tau), u(t)), \quad \tau \in [t, t+T],
\end{align}
with the initial condition $\Tilde{x}(t) = x(t)$.
Then we can define the estimator from $t$ to $t+T$ as
\begin{align}
g(x(t),u(t)) := h(\Tilde{x}(t+T)). 
\end{align}
The Newton-Raphson controller proposed in \cite{Wardidetail} has the following form:
\begin{equation}
    \dot u(t) = \alpha \left ( \frac{d g }{d u}(x(t), u(t))\right )^{-1}(r(t+T)-g(x(t),u(t)))
\end{equation}
where $r(t+T)$ is assumed to be known in advance at time $t$.
Please note that whether the system is actuated or underactuated will not influence the work of the controller but the system should be controllable.

The calculation of $\frac{d g(x(t), u(t))}{d u}$, however, has a high computation demand if the system is nonlinear.
There are three ways to calculate $\frac{d g(x(t), u(t))}{d u}$, to the authors' best knowledge:
\begin{enumerate}
    \item Finite difference method (FDM) where 
    \begin{align*}
    \frac{dg(x(t),u(t))}{du}=\frac{g(x,u+\delta u)-g(x,u)}{\delta u}.
    \end{align*}
    This method is direct but very time-consuming.
    \item Without loss of generality, assume $h(\Tilde{x}(t+T))=\Tilde{x}(t+T)$. Use Eqn. \eqref{eqn:deri2}:
    \begin{align} \label{eqn:goal3}
    \frac{dg(x(t),u(t))}{du(t)}=\frac{d\Tilde{x}(t+T)}{du(t)}
    \end{align}
    where
    \begin{align} \label{eqn:m3}
    \overset{\cdot}{\left [  \frac{d\Tilde{x}(\tau)}{du(t)}   \right ]}= \frac{\partial f(\Tilde{x}(\tau),u(t))}{\partial x} \frac{d\Tilde{x}(\tau)}{du(t)}+\frac{\partial f(\Tilde{x}(\tau),u(t))}{\partial u}.
    \end{align}
    By defining a new variable $\xi(t)=\frac{d\Tilde{x}(\tau)}{du(t)}$, we can obtain the \eqref{eqn:goal3} by solving ODE \eqref{eqn:m3}.
    \item The third method is based on linearized models ($\dot x=Ax+Bu$ and $y=Cx$). 
    If the model is linear or we linearize the nonlinear system locally, we have the predicted output
    \begin{align}
      y(t+T) = C_t(e^{A_t T}x_t + A_t^{-1}(e^{A_t T}-I_n)(B_t u)).
    \end{align}
    where $I_n$ is the $n\times n$ identity matrix.
  Thus,
   \begin{align}
    \frac{\partial y(t+T)}{\partial u} = C_tA_t^{-1}(e^{A_t T}-I_n)B_t,
\end{align}
This method, however, only works for linear models. If a model is nonlinear and we linearized locally, it is possible that the linearized model is not controllable which makes this method not feasible in this case.
For example, the Dubins car is controllable but the linearized model of the Dubins car is not controllable.
\end{enumerate}
Therefore, we propose using Koopman operator theory to lift the nonlinear systems to linear systems, thus alleviating the complexity and still keeping the controllability of the original nonlinear systems.
\section{Newton-Raphson controller based on Koopman Operator theory} \label{sec:pc}
In this section, we will first give a brief introduction to the principle of Koopman operator.
Based on this, we propose our controller based on the linearized model obtained by Koopman operator theory.
\subsection{Koopman operator theory \cite{soft}}
The Koopman operator provides a linear representation of the flow of a nonlinear system in an infinite-dimensional space of observables.
Consider a dynamical system
\begin{align*}
\dot x = F(x(t))
\end{align*}
where $x(t)\in \mathbb{R}^n$ and $F$ is a continuously differentiable function.
The system can be lifted to an infinite-dimensional function space $\mathcal{F}$ composed of all continuous real-valued functions.
In $\mathcal{F}$, the flow of the system is characterized by the linear Koopman operator $U_t:\mathcal{F}\rightarrow \mathcal{F}$ which describes the evolution of the observables along the trajectories of the system.
We seek the projection of the Koopman operator onto a finite-dimensional subspace.
Denote $\Bar{\mathcal{F}}\subset \mathcal{F}$ to be the subspace of $\mathcal{F}$ spanned by $N>n$ linearly independent basis function $\{\phi_i:\mathbb{R}^n \rightarrow \mathbb{R} \}_{i=1}^N$.
For convenience, we assume the first $n$ basis functions are the states, i.e., $\phi_i(x)=x_i$.
Thus, written as a vector form, we have
\begin{align}
\phi(x) = [x_1, x_2,\cdots x_n,\phi_{n+1}(x), \cdots, \phi_N(x)].
\end{align}
Any observables $\bar{f}\in \bar{\mathcal{F}}$ can be expressed as a linear combination of elements of these basis functions, i.e.,
\begin{align*}
\bar{f} = w_1 \phi_1+w_2\phi_2+\cdots +w_N \phi_N = w^\top \phi(x)^\top
\end{align*}
where $w_i\in \mathbb{R}$.
The $\phi(x)$ is called the lifted state and $w$ is the vector representation of $\bar{f}$.
Given this representation, we can obtain an approximation of the Koopman operator $U_t\in \mathbb{R}^{N\times N}$ on $\mathcal{F}$ that satisfies
\begin{align*}
\bar{U}_t w = w'
\end{align*}
The objective is to find the $\bar{U}_t$ based on observable data in $\bar{\mathcal{F}}$.

\subsection{Proposed controller design}
For dynamical systems with inputs Eqn. \eqref{eqn:system}, we aim to build a linear model from the Koopman operator theory aforementioned:
\begin{align} \label{eqn:liftedsystem}
z[(j+1)] &= A z[j]+B u[j] \\
x[j]   &=  C z[j]
\end{align}
for each $j \in \mathbb{N}$, $A\in \mathbb{R}^{N\times N}$ is the state transition matrix, $z=\phi(x)$ is the lifted state, $B\in \mathbb{R}^{N \times m}$ is the control matrix, and $C=\begin{bmatrix}
I_{n\times n} &0_{n\times (N-n)} 
\end{bmatrix}$ is a projection operator from the lifted space onto the state space.

Denote 
\begin{align*}
\alpha[k] = \left [  x[k], u[k] \right ]\\
\beta[k] = \left [ F(x[k],u[k]),u[k] \right ].
\end{align*}
We then identify a finite-dimensional approximation of the Koopman operator via the Extending Dynamic Mode Decomposition (EDMD) algorithm \cite{EDMD} using observed data.
The corresponding Koopman operator is
\begin{align*}
\bar{U}_{T_s}=\Gamma_c^\dagger
\Gamma_n,
\end{align*}
where $\dagger$ means pseudo-inverse, $K$ is the time horizon for collecting data, $\beta[k]=F(\alpha[k]), \;k=1,2,\cdots K$, and
\begin{align*}
\Gamma_c = \frac{1}{ K}\sum_{k=1}^{K} \phi(\alpha[k])^\top\phi(\alpha[k]), \\
\Gamma_n = \frac{1}{ K}\sum_{k=1}^{K} \phi(\alpha[k])^\top\phi(\beta[k])
\end{align*}
The continuous Koopman operator can then be written as $\log(\bar{U}_{T_s})/\Delta t$ where $\Delta t$ is the sampling time.
The $\bar{U}_{T_s}^\top$ is the best approximation of a transition matrix between the elements of snapshot pairs in the $L^2$-norm sense, i.e.,
\begin{align}
\min_{U_{T_s}^\top} \sum_{k=1}^K  \left \|U_{T_s}^\top \phi(\alpha[k])-\phi(\beta[k]) \right \|_2^2.
\end{align}
The best $A$ and $B$ matrices in \eqref{eqn:liftedsystem} can be isolated by partitioning the $\bar{U}_{T_s}^\top$:
\begin{align*}
\bar{U}_{T_s}^\top = \begin{bmatrix}
A_{N\times N} & B_{N\times m}\\ 
0_{m\times N} & I_{m\times m}
\end{bmatrix},
\end{align*}
where $I_{m\times m}$ is the identity matrix.

\begin{figure}[!htp]
    \centering    \includegraphics[width=\linewidth]{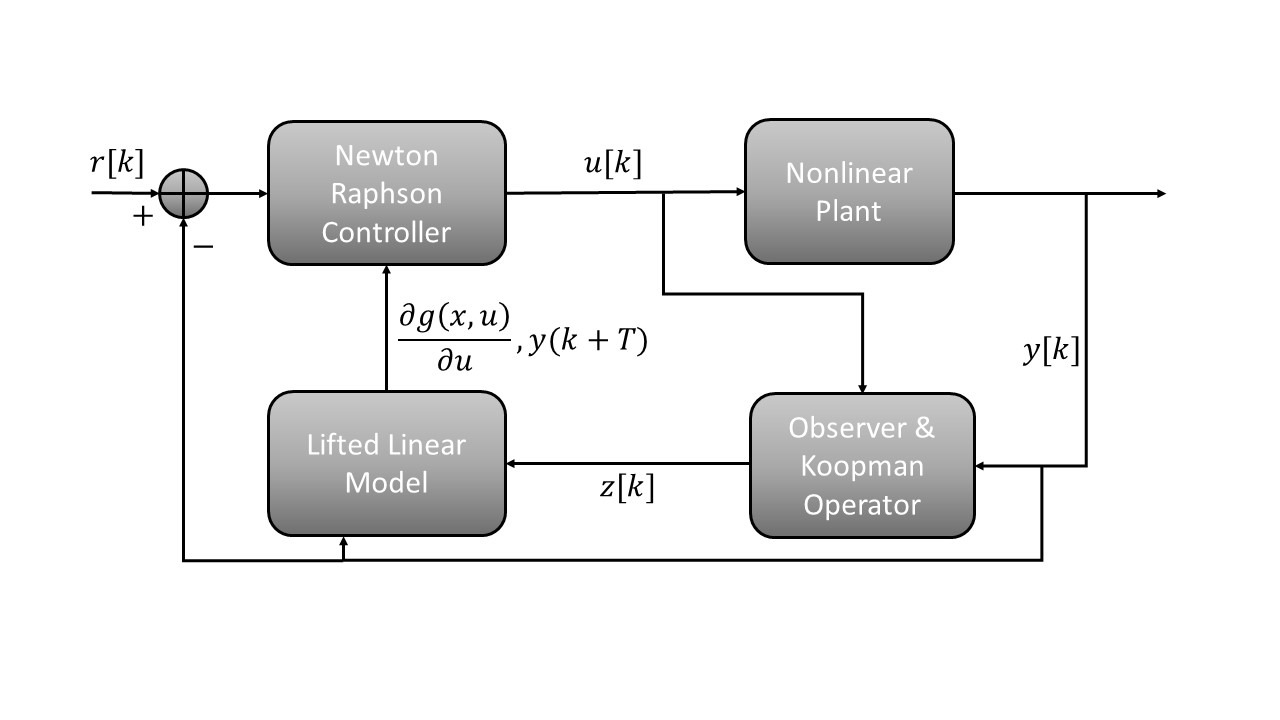}
    \caption{Diagram of the proposed control framework: A lift linear system are built based on Koopman operator theory; the derivative and prediction of Newton-Raphson controller is calculated using the linearized model.}
    \label{fig:diagram}
    \vspace{-10pt}
\end{figure}

Fig. \ref{fig:diagram} shows the diagram of the proposed data-driven Newton-Raphson tracking scheme.
In this algorithm, we first collect data from the nonlinear system and build a lifted linear system from the collected data.
After that, the predictor $y(t+T)$ and the derivative term $\frac{\partial g(x,u)}{\partial u}$ are obtained from the linearized model.
In this way, we avoid the problems aforementioned.
\section{Simulation} \label{sec:exp}
In this section, we provide three examples to verify the efficiency of the proposed controller.
We then compare the results with that of the classical Newton-Raphson controller with respect to tracking accuracy and time.
All the experiments are implemented on a personal computer with MATLAB R2020b.
For all the experiments, we use mean square error as the measure for tracking performance, which is defined as
\begin{align*}
MSE = \frac{1}{N_d}\sum_{i=1}^{N_d} ||y(t_i)-r(t_i)||_2^2,
\end{align*}
where $N_d=\frac{t_f}{dt}$ is the number of sampling points.
The average MSE and time complexity of 10 experiments are taken for comparison.
\subsection{Example 1: Van Der Pol system}
A typical Van Der Pol system has the following form:
\begin{align}
\left\{\begin{matrix}
\dot x_1 = x_2 \\
\dot x_2 = -x_1 + (1-x_1^2)x_2+u
\end{matrix}\right..
\end{align}
The objective is to make $y=x_1$ tracking the signal $r(t)= \frac{\pi}{8}\sin t +\frac{\pi}{6}$.
The basis is chosen as $z=[x_1, x_2, x_1^2, x_1^2x_2]^\top$ \footnote{Please note that this choice of basis is not unique.}.
Using EDMD algorithm, we obtain constant matrix $A_{4\times 4},\; B_{4 \times 1}$, with which we rebuild a linearized system $\dot z = Az+Bu$ from the collected data.
The derivative $\frac{d g(x,u)}{d u}$ and the predictor $y=x_1$ can then be obtained by using this linearized system.
For the identification using Koopman operator theory, the prediction horizon is chosen as $T_s = 2 \mathrm{s}$ and the numbers for trials for data collection is $N_s = 10$ with random initialization of the initial state and control input \footnote{Please note that the larger the value of $N_h$ and $N_t$, the better the tracking results.
However, to compromise the time complexity, we chose this group of parameters.}.
The initial state is set as $[0,0]^\top$.
The speedup parameter is chosen as $\alpha=20$ for both the NR controller and KNR controller.
The sampling time is $0.01 \; \mathrm{s}$.
The prediction horizon is $T=0.15 \; \mathrm{s}$.
The system is simulated for $t_f=20 \mathrm{s}$.

\begin{figure}[!htp]
    \centering
   \includegraphics[width=\linewidth]{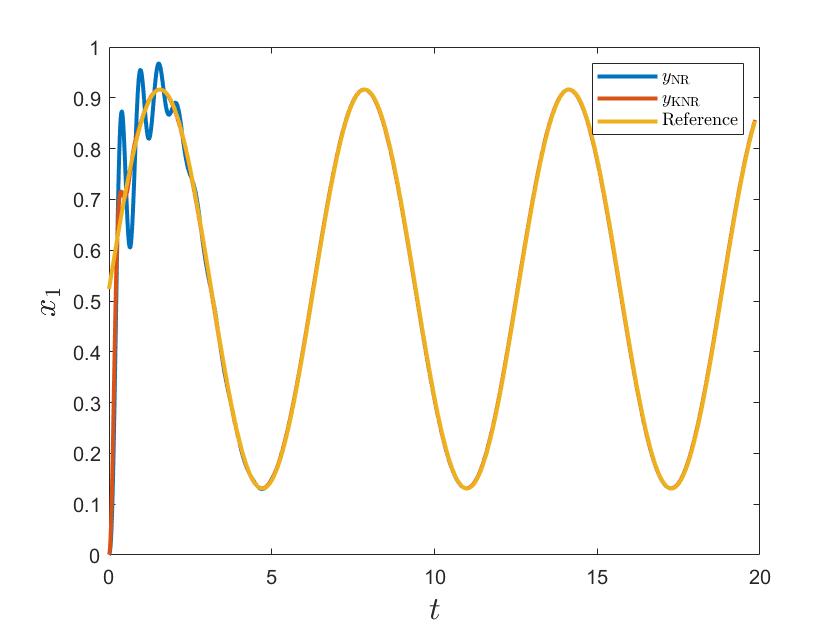}
    \caption{Tracking results for the Van Der Pol system.}
    \label{fig:vdp_traj}
\end{figure}

\begin{figure}[!htp]
    \centering
   \includegraphics[width=\linewidth]{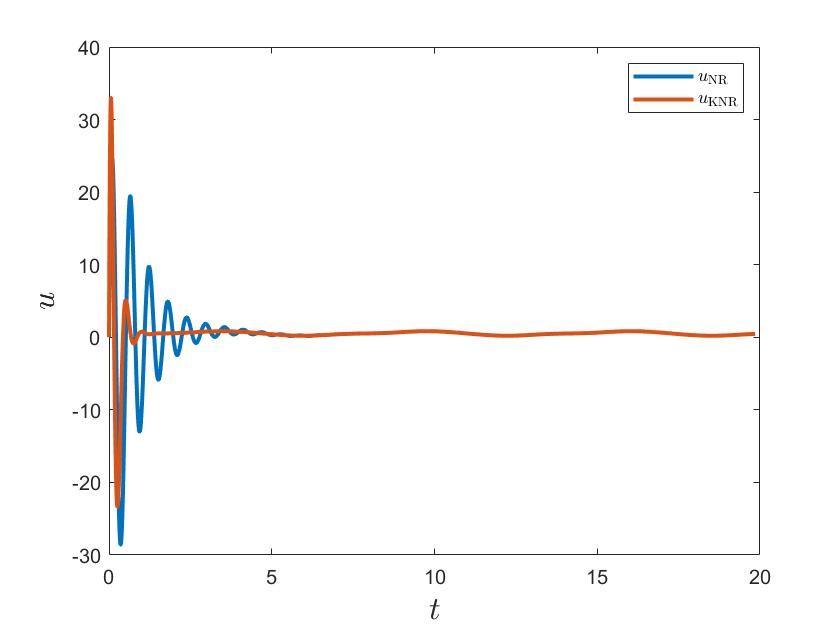}
    \caption{Control input.}
    \label{fig:vdp_control}
\end{figure}

Fig. \ref{fig:vdp_traj} is the trajectory of $x_1(t)$ for both KNR and NR.
As we can see, KNR has a smaller oscillation at the beginning than that of NR.
Fig. \ref{fig:vdp_control} is the control input by using both algorithms.
It shows that the KNR reaches zero faster than the NR and has fewer oscillations as well.

Table \ref{tab:VanderPol} summarizes the compared results with the classical Newton-Raphson (NR) controller.
All the parameters keep the same for both algorithms.
As we can see, the KNR needs less time to catch up with the reference signal and has higher accuracy.
This is easy to be explained since when calculating the $\frac{d g(x,u)}{d u}$, the linearized system has higher accuracy.
Regarding the time complexity of KNR in Table \ref{tab:VanderPol}, the time can be reduced on a large scale by reducing the number of trials and size of the prediction window.
Thus the difference of time complexity between NR and KNR may be negligible.
\begin{table}[!htp]
    \centering
    \begin{tabular}{||c|c|c||} \hline
       Algorithm  & Mean squared error (MSE) & Time [$\mathrm{s}$]\\ \hline
       NR  & 0.0027 & 1.1416 \\ \hline
       KNR & 0.0017 & 1.2769 \\ \hline
    \end{tabular}
    \caption{Compare results for Van Der Pol system.}
    \label{tab:VanderPol}
    \vspace{-20pt}
\end{table}
\subsection{Example 2: Overhead crane system}
\begin{figure}[!htp]
    \centering   \includegraphics[width=\linewidth]{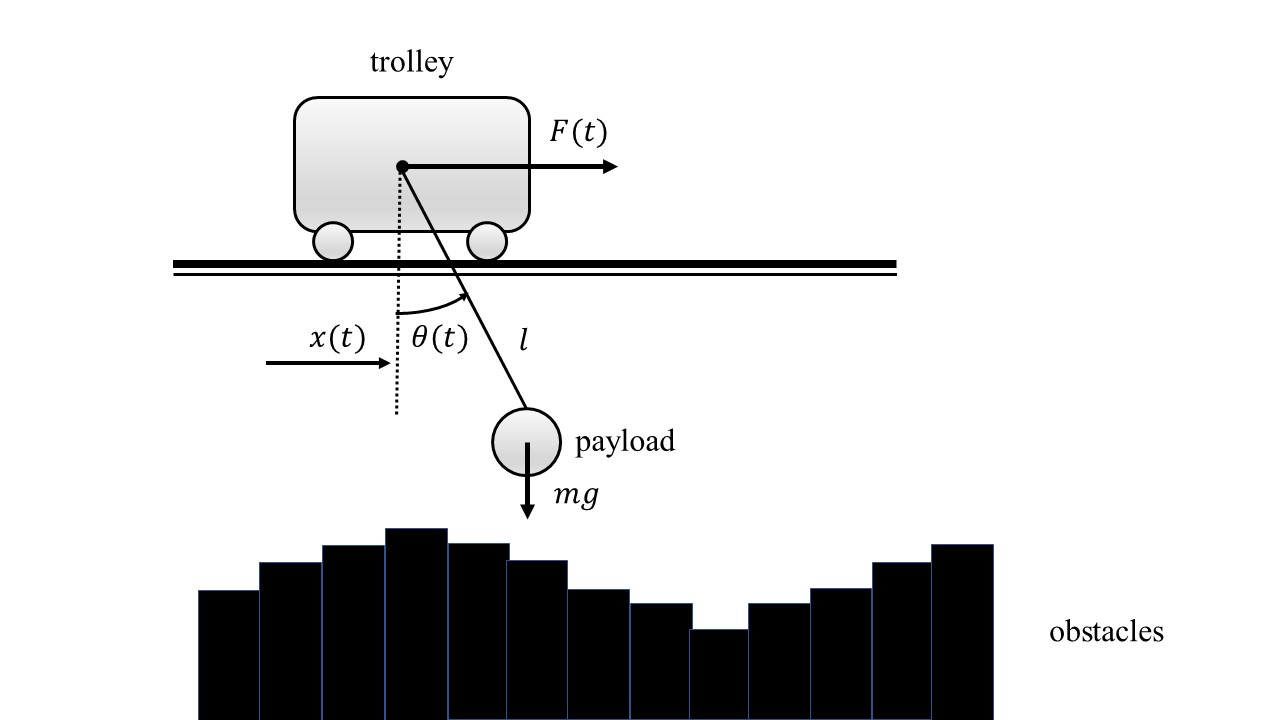}
    \caption{Illustration for a planar two-dimensional overhead crane system tracking a pre-defined trajectory.}
    \label{fig:crane}
\end{figure}

\begin{figure}[!htp]
    \centering
    \includegraphics[width=\linewidth]{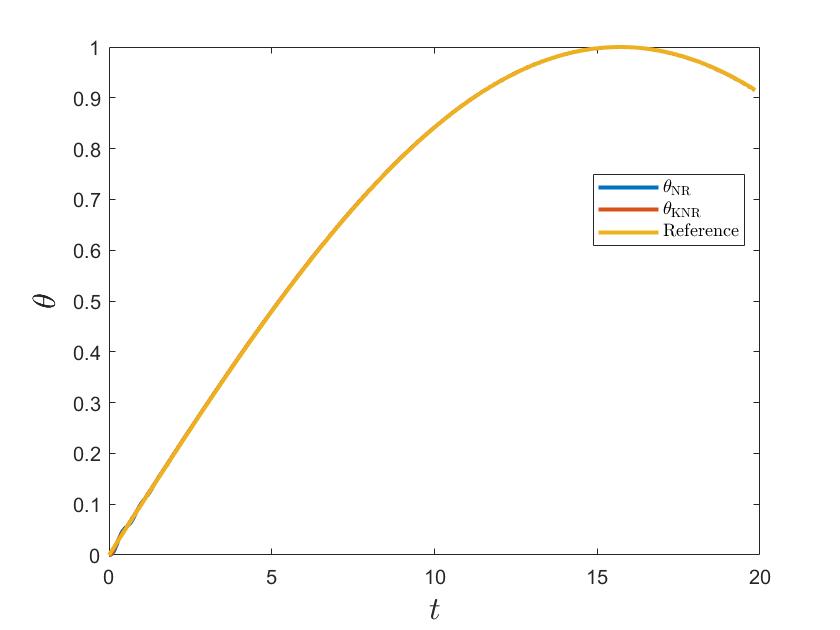}
    \caption{Tracking results for the overhead crane system.}
    \label{fig:crane_track}
\end{figure}

\begin{figure}[!htp]
    \centering
    \includegraphics[width=\linewidth]{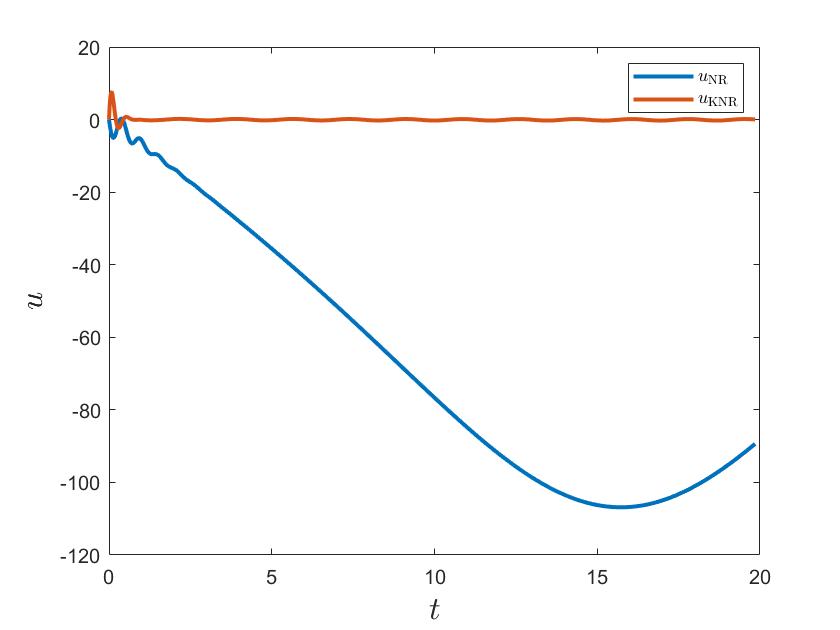}
    \caption{Control input (i.e., $F(t)$) of the overhead crane system.}
    \label{fig:crane_input}
\end{figure}

The overhead crane system has the following dynamics \cite{Mi-Nankai}:
\begin{align}
(M+m) \Ddot{x} + m l \Ddot{\theta} \cos \theta-ml\dot \theta^2\sin \theta = F \nonumber\\
ml^2 \Ddot{\theta} + ml \cos\theta \Ddot{x} +mgl \sin\theta = 0
\end{align}
where $M$ and $m$ denote the masses of the trolley and payload respectively;
$l$ is the length of the rod (constant);
$g$ is the acceleration of gravity ($9.81\; \mathrm{m/s^2}$).
$x(t)$ is the position of the trolley actuated by the control torque $F(t)$ while $\theta(t)$ represents the underactuated payload swing angle.
Suppose there is an obstacle and we need to transport the payload by tracking a defined trajectory $r(t)$.
Fig. \ref{fig:crane} shows the defined problem where we regard the contour of the obstacles as some predefined tracking trajectory to avoid the obstacles.
Same as \cite{Mi-Nankai}, the physical parameters of the crane test bed are taken as $M=6.5 \mathrm{kg}$, $m = 0.5 \mathrm{kg}$, $l=0.75 \mathrm{m}$.
The initial state is $[0,0,0,0]^\top$.
The objective is to let the observation $\theta$ track a predefined signal which is the contour of the obstacles.
Let the predefined signal $r(t)=\sin(0.1 t)$ and the simulation time $t_f=20$.
The prediction horizon is $T=0.15$.
The speedup parameter is $\alpha=20$.
The basis for the lifted model is chosen as $[x,\dot x, \theta, \dot \theta, \sin(\theta),\cos(\theta)]$.
Fig. \ref{fig:crane_track} is the tracking result of the proposed controller and Fig. \ref{fig:crane_input} is the corresponding control input.
As we can see, both KNR and NR can track the reference signal very well but the control input of KNR has less magnitude.

Table \ref{tab:crane} summarizes the compare results for the overhead crane system.
\begin{table}[!htp]
    \centering
    \begin{tabular}{||c|c|c||} \hline
       Algorithm  & Mean squared error (MSE) & Time [$\mathrm{s}$] \\ \hline
       NR  &7.3090e-07  & 1.1560 \\ \hline
       KNR & 5.8774e-07 & 1.3684  \\ \hline
    \end{tabular}
    \caption{Compare results for overhead crane system.}
    \label{tab:crane}
    \vspace{-20pt}
\end{table}

\subsection{Example 3: Differentially driven car}
\begin{figure}[!htp]
    \centering
    \includegraphics[width=\linewidth]{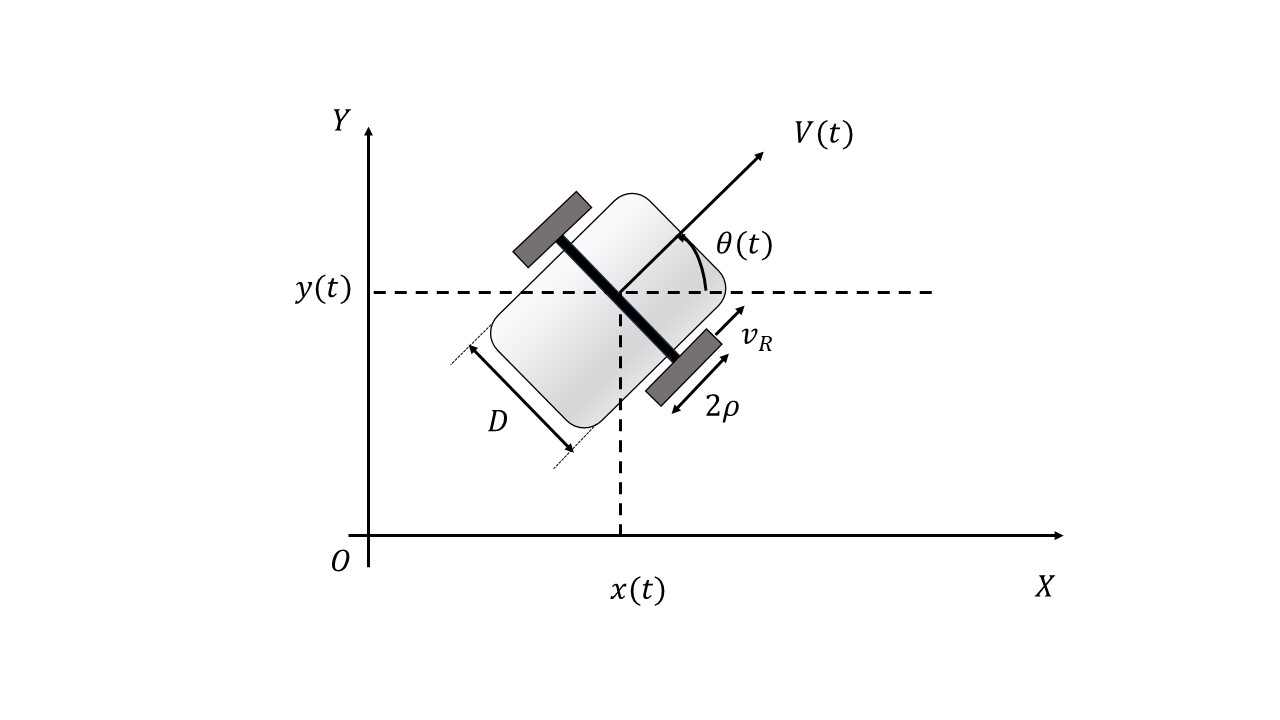}
    \caption{Illustration of a differentially driven car.}
    \label{fig:car}
\end{figure}

\begin{figure}[!htp]
    \centering
    \includegraphics[width=\linewidth]{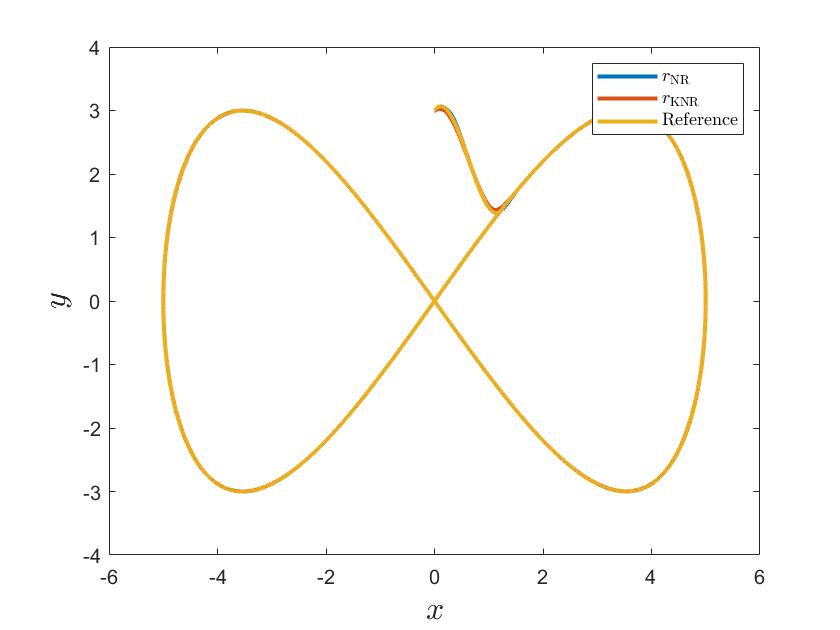}
    \caption{Tracking result for the differentially driven car.}
    \label{fig:car_track}
\end{figure}

\begin{figure}[!htp]
    \centering
    \includegraphics[width=\linewidth]{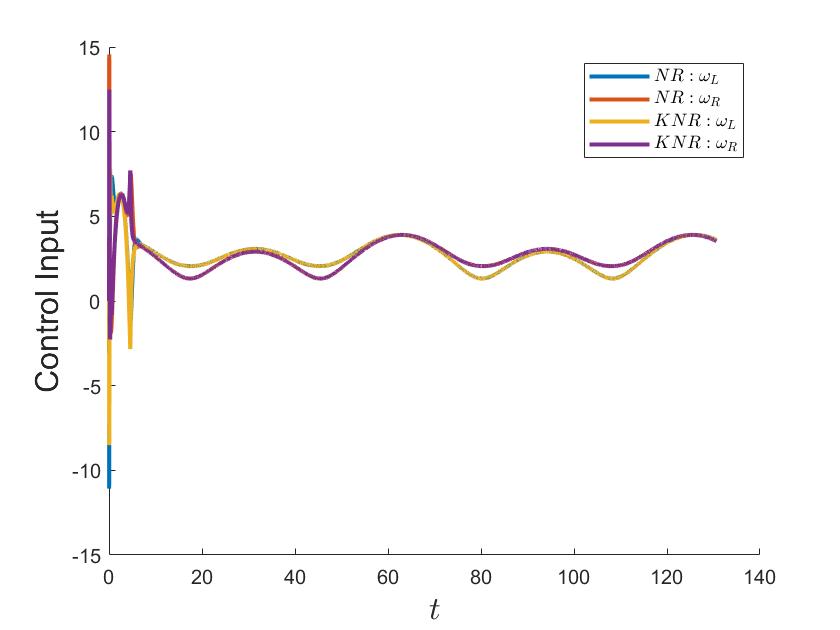}
    \caption{Control input for the differentially driven car.}
    \label{fig:car_input}
\end{figure}

The vehicle's kinematic dynamic in global coordinates is as follows \cite{Niu}:
\begin{align}
\begin{bmatrix}
\dot x (t)\\ 
\dot y(t)\\ 
\dot \theta(t)
\end{bmatrix} = \begin{bmatrix}
\frac{\rho}{2}\cos\theta(t) &  \frac{\rho}{2}\cos\theta(t)\\ 
\frac{\rho}{2}\sin \theta(t) & \frac{\rho}{2}\sin \theta(t)  \\ 
-\frac{\rho}{D} & \frac{\rho}{D}  
\end{bmatrix}\begin{bmatrix}
\omega_L(t)\\ 
\omega_R(t)
\end{bmatrix}
\end{align}
where the system state is $(x(t), y(t), \theta(t))^\top$ and the control input is $u(t) = (\omega_L(t), \omega_R(t))^\top$.
Physically, $x(t)$ (resp. $y(t)$) is the $x$ (resp. $y$) position of the car in the world coordinates.
$\theta(t)$ is the orientation of the car with respect to the global coordinates system as shown in Fig. \ref{fig:car}.
$\omega_R(t) = v_R/\rho$ (resp. $\omega_L(t)=v_L/\rho$) is the angular velocity of the right wheel (resp. left wheel).
$\rho$ is the radius of the wheels.
$D$ is the width of the vehicle.
As we can see, this system is highly nonlinear.
Let $\rho=0.1 \; \mathrm{m}$, $D=0.4\; \mathrm{m}$.
Similar to \cite{Niu}, we define the following reference trajectory $r(t)$:
\begin{align}
r(t) =
\left\{\begin{matrix}
(-0.0001t^3+0.25t, 0.0475t^3-0.3601t^2+0.3t+3),\\ t<5 \\ 
(5 \sin (0.05 t),\; 3 \sin (0.1t)), \; t>5
\end{matrix}\right. .
\end{align}
Let the initial position be $(x_1(0), x_2(0))^\top = (0,3)^\top$ and the initial angle be $\theta=0 \;\mathrm{rad}$.
The total simulation time is $t_f=130.6637$; the look-ahead time is $T=0.5\; \mathrm{s}$ and the time step is $dt=0.01$.
The speedup parameter is chosen as $\alpha=20$ for both the NR controller and KNR controller.
The basis is chosen as $z=[x,y,\theta,\omega_L\sin\theta,\omega_L\cos\theta,\omega_R\sin\theta,\omega_R\cos\theta]$.
In this example, to compute $\frac{dg(x,u)}{du}$, we use Method 2 as listed in Section \ref{sec:PS}.
This method is more efficient than the finite difference method (FDM) as used in Example 1 and Example 2.
%By using Method 2 directly in the linearized model and setting $\xi(t)=[0, 0; 0, 0; 0,0; \sin(\theta(t)),0; \cos(\theta(t)),0; 0, \sin(\theta(t));0, \cos(\theta(t))]$, we obtain the derivative $\frac{dg(x,u)}{du}$.
%Method 2 is used for both NR and KNR.
Fig. \ref{fig:car_track} is the trajectory given by the NR controller ($r_{\mathrm{NR}}$), KNR controller ($r_{\mathrm{KNR}}$), and the reference signal.
Fig. \ref{fig:car_input} is the corresponding control input.
As we can see, KNR has almost the same tracking performance as NR.
What's more, the control input has less magnitude.

Table \ref{tab:car} summarizes the compare results for the differentially driven car system.
In this example, the time of KNR is longer than that of NR due to the system identification process using the Koopman theory.
\begin{table}[!htp]
    \centering
    \begin{tabular}{||c|c|c||} \hline
       Algorithm  & Mean squared error (MSE) & Time [$\mathrm{s}$]\\ \hline
       NR  &1.4395e-05 &  0.8774 \\ \hline
       KNR &3.2003e-05 &  0.9095 \\ \hline
    \end{tabular}
    \caption{Compare results for differentially driven car.}
    \label{tab:car}
    \vspace{-20pt}
\end{table}
\section{Conclusion} \label{sec:conclude}
This work introduced the utilization of the Koopman operator theory to achieve a model-free Newton-Raphson controller.
Three highly nonlinear systems are provided to verify our method.
One possible extension is to reduce the overshoot of the controller in the beginning.
In the future, we expect to implement some experiments in the Robotarium platform and more complicated nonlinear systems.

\bibliographystyle{./IEEEtran} 
\bibliography{./IEEEabrv,./IEEEexample}

\begin{thebibliography}{10}
\providecommand{\url}[1]{#1}
\csname url@rmstyle\endcsname
\providecommand{\newblock}{\relax}
\providecommand{\bibinfo}[2]{#2}
\providecommand\BIBentrySTDinterwordspacing{\spaceskip=0pt\relax}
\providecommand\BIBentryALTinterwordstretchfactor{4}
\providecommand\BIBentryALTinterwordspacing{\spaceskip=\fontdimen2\font plus
\BIBentryALTinterwordstretchfactor\fontdimen3\font minus
  \fontdimen4\font\relax}
\providecommand\BIBforeignlanguage[2]{{%
\expandafter\ifx\csname l@#1\endcsname\relax
\typeout{** WARNING: IEEEtran.bst: No hyphenation pattern has been}%
\typeout{** loaded for the language `#1'. Using the pattern for}%
\typeout{** the default language instead.}%
\else
\language=\csname l@#1\endcsname
\fi
#2}}

\bibitem{leaderfor}
\BIBentryALTinterwordspacing
Y.~Wardi, C.~Seatzu, and M.~Egerstedt, ``Tracking control via variable-gain
  integrator and lookahead simulation: Application to leader-follower
  multiagent networks,'' \emph{IFAC-PapersOnLine}, vol.~51, no.~16, pp.
  217--222, 2018, 6th IFAC Conference on Analysis and Design of Hybrid Systems
  ADHS 2018. [Online]. Available:
  \url{https://www.sciencedirect.com/science/article/pii/S2405896318311534}
\BIBentrySTDinterwordspacing

\bibitem{selfdriving}
T.~Faulwasser, B.~Kern, and R.~Findeisen, ``Model predictive path-following for
  constrained nonlinear systems,'' 12 2009, pp. 8642--8647.

\bibitem{UAV}
H.~Lee, ``Trajectory tracking control of multirotors from modelling to
  experiments: A survey,'' \emph{International Journal of Control, Automation
  and Systems}, vol.~15, 12 2016.

\bibitem{underwater}
\BIBentryALTinterwordspacing
Z.~Bingul and K.~Gul, ``Intelligent-pid with pd feedforward trajectory tracking
  control of an autonomous underwater vehicle,'' \emph{Machines}, vol.~11,
  no.~2, 2023. [Online]. Available:
  \url{https://www.mdpi.com/2075-1702/11/2/300}
\BIBentrySTDinterwordspacing

\bibitem{regulator}
A.~Isidori and C.~Byrnes, ``Output regulation of nonlinear systems,''
  \emph{IEEE Transactions on Automatic Control}, vol.~35, no.~2, pp. 131--140,
  1990.

\bibitem{mpcbook}
E.~F. Camacho and C.~B. Alba, \emph{Model predictive control}.\hskip 1em plus
  0.5em minus 0.4em\relax Springer science \& business media, 2013.

\bibitem{MPC}
\BIBentryALTinterwordspacing
J.~Bettega and D.~Richiedei, ``Trajectory tracking in an underactuated,
  non-minimum phase two-link multibody system through model predictive control
  with embedded reference dynamics,'' \emph{Mechanism and Machine Theory}, vol.
  180, p. 105165, 2023. [Online]. Available:
  \url{https://www.sciencedirect.com/science/article/pii/S0094114X22004104}
\BIBentrySTDinterwordspacing

\bibitem{Wardi}
Y.~Wardi, C.~Seatzu, M.~Egerstedt, and I.~Buckley, ``Performance regulation and
  tracking via lookahead simulation: Preliminary results and validation,'' in
  \emph{2017 IEEE 56th Annual Conference on Decision and Control (CDC)}, 2017,
  pp. 6462--6468.

\bibitem{Wardidetail}
Y.~Wardi, C.~Seatzu, J.~Cort{\'e}s, M.~Egerstedt, S.~Shivam, and I.~Buckley,
  ``Tracking control by the newton-raphson method with output prediction and
  controller speedup,'' \emph{arXiv: Optimization and Control}, 2019.

\bibitem{formation}
Y.~Wardi, C.~Seatzu, and S.~Shivam, ``Application of an output tracking
  technique to formation control,'' in \emph{2021 29th Mediterranean Conference
  on Control and Automation (MED)}, 2021, pp. 898--903.

\bibitem{TA}
S.~Shivam, A.~Kanellopoulos, K.~G. Vamvoudakis, and Y.~Wardi, ``A predictive
  deep learning approach to output regulation: The case of collaborative
  pursuit evasion,'' in \emph{2019 IEEE 58th Conference on Decision and Control
  (CDC)}, 2019, pp. 853--859.

\bibitem{Niu}
K.~Niu, Y.~Wardi, C.~T. Abdallah, and M.~Hayajneh, ``A model-free tracking
  controller based on the newton-raphson method and feedforward neural
  networks,'' in \emph{2022 American Control Conference (ACC)}, 2022, pp.
  3254--3259.

\bibitem{Brunton2019NotesOK}
S.~L. Brunton, ``Notes on koopman operator theory,'' 2019.

\bibitem{soft}
D.~Bruder, X.~Fu, B.~Gillespie, C.~Remy, and R.~Vasudevan, ``Koopman-based
  control of a soft continuum manipulator under variable loading conditions,''
  \emph{IEEE Robotics and Automation Letters}, vol.~PP, pp. 1--1, 07 2021.

\bibitem{DualModeMPC}
A.~Krolicki, D.~Tellez-Castro, and U.~Vaidya, ``Nonlinear dual-mode model
  predictive control using koopman eigenfunctions,'' in \emph{2022 IEEE 61st
  Conference on Decision and Control (CDC)}, 2022, pp. 3074--3079.

\bibitem{aerospace}
S.~Servadio, R.~Armellin, and R.~Linares, ``A koopman-operator control
  optimization for relative motion in space,'' 2022.

\bibitem{EDMD}
M.~Williams, I.~Kevrekidis, and C.~Rowley, ``A data-driven approximation of the
  koopman operator: Extending dynamic mode decomposition,'' \emph{Journal of
  Nonlinear Science}, vol.~25, 08 2014.

\bibitem{Mi-Nankai}
M.~Zhou, N.~Sun, H.~Chen, and Y.~Fang, ``A novel sliding mode control method
  for underactuated overhead cranes,'' in \emph{2017 Chinese Automation
  Congress (CAC)}, 2017, pp. 4548--4552.

\end{thebibliography}

\end{document}